\documentclass[lettersize,journal]{IEEEtran}
\usepackage{amsmath,amsfonts}
\usepackage{algorithmic}
\usepackage{algorithm}
\usepackage{array}
\usepackage[caption=false,font=footnotesize,labelfont=normalfont,textfont=normalfont]{subfig}
\usepackage{textcomp}
\usepackage{url}
\usepackage{verbatim}
\usepackage{cite}
\usepackage{setspace}
\usepackage{amsmath,amsfonts,amsthm}
\usepackage{graphicx}      
\usepackage[dvipsnames]{xcolor}
\usepackage{float}
\usepackage{hyperref}

 \theoremstyle{definition}
\if@secthm
  \newtheorem{thm}{Theorem}[section]
  \@addtoreset{thm}{section}
\else
  \newtheorem{thm}{Theorem}
\fi
\newtheorem{cor}[thm]{Corollary}

\newtheorem{assum}[thm]{Assumption}
\newtheorem{prop}[thm]{Proposition}
 \theoremstyle{definition}

\newtheorem{rem}[thm]{Remark}

\usepackage[prependcaption,colorinlistoftodos]{todonotes}

\begin{document}

\title{Frequency Control and {Optimal} Power Sharing\\ in Combined Power and Heating Networks \\ {with Heat Pumps}}

\author{Xin Qin 
        and Ioannis Lestas 
\thanks{X. Qin and I. Lestas are with the Department of Engineering, University of                  Cambridge, Cambridge, CB2 1PZ, UK
        {\tt\small \{xq234, icl20\}@ cam.ac.uk}. {Conference paper \cite{qin2024frequency} is a preliminary version of this work. This manuscript is a significantly longer document that includes the derivations of the results, extended simulation case studies and additional analysis and discussion.}}
}



\maketitle

\begin{abstract}
Heat pumps have the capability for fast adjustments in power consumption with potential connections to large heating-inertia district heating networks, and are thus a very important resource for providing frequency support in low-inertia power systems. Nevertheless, the coupling of power networks with district heating systems renders the underlying dynamics much more involved. It is therefore important to ensure that system stability and appropriate power sharing are maintained. In this paper, we consider the problem of leveraging district heating systems as ancillary services for primary frequency control in power networks via heat pumps.
We propose a novel power sharing scheme for heating systems based on the average temperature.
This enables an optimal power allocation among diverse energy sources without {requiring load disturbances information}.
We then discuss two approaches for heating systems to contribute to frequency regulation in power networks. We show that both approaches ensure stability in the combined heat and power network and facilitate optimal power allocation among the different energy sources.
We also discuss how various generation dynamics can be incorporated into our framework with guaranteed stability and optimality.
Finally, we conduct simulations that demonstrate various tradeoffs in the transient response and the practical potential of the proposed approaches.
\end{abstract}



\section*{Nomenclature}

\addcontentsline{toc}{section}{Nomenclature}
\subsection{Frequently Used Parameters}
\begin{IEEEdescription}[\IEEEusemathlabelsep\IEEEsetlabelwidth{$B_{k,i}$}]
\begin{spacing}{1.15}
\item[$M_j$]  Generator inertia {at} bus $j$
\item[$D_j$]  Damping coefficient {at} bus $j$

\item[$V^E_j$]  Volume of heat edge $j$
\item[$V^N_j$]  Volume of heat node $j$
\item[$p_j^L$]  Electric load demand deviation {at} bus $j$
\item[$h^L_j$]  Heat load demand deviation {at} edge $j$
\item[$q^E_j$]  {Mass flow at} edge $j$
\item[$C_{o,j}$]  Coefficient of performance of {the} heat pump {at} bus~$j$
\item[$Q_{e,jj}$] Cost coefficient of electric generators {at} bus $j$
\item[$Q_{h,jj}$] Cost coefficient of conventional heat source {at} edge~$j$
\item[$Q_{u,jj}$] Cost coefficient of {power imbalance at} {bus} $j$
\end{spacing}
\end{IEEEdescription}

\subsection{Frequently Used Variables}
\begin{IEEEdescription}[\IEEEusemathlabelsep\IEEEsetlabelwidth{$B_{k,i}$}]
\begin{spacing}{1.1}
\item [$\omega_{j}$] Frequency deviation {at} bus $j$ from a nominal value
\item[$\eta_{ij}$]  Angle difference between bus $i$ and $j$
\item[{$p_{ij}$}]  {Power transfer between bus $i$ and $j$}
\item[$p^G_j$]  Deviation in electric power generation by {the} generator {at} bus $j$
\item[$p^P_j$]  Deviation in electric power consumption of the heat pump {at} bus $j$
\item[$h^G_{j}$] Deviation in heat power generation of conventional heat source {at} edge $j$
\item[$h^P_{j}$]  Deviation in heat power generation of heat pump power {at} edge $j$
\end{spacing}
\end{IEEEdescription}

\section{Introduction}

\subsection{Background and motivation}

As decarbonization efforts intensify and natural gas prices increase, power systems are experiencing an increasing penetration of renewable generation and the retirement of fossil-fuel generators. In this transition, the uncertainty and low inertia renewable generation require more resources providing support to frequency regulation.
As ancillary markets become saturated by battery storage and less profitable~\cite{ref_saturation}, it is urgent to find flexible and reliable resources for frequency regulation with affordable costs.

Meanwhile, on the demand side, the shift {towards} electrified heating highlights the growing importance and potential of heat pumps in {supporting frequency regulation}~\cite{ref_ukgov_hp}.
{Compared to traditional electric loads and heat sources, heat pumps not only have
rapid response capabilities but also interlink two different energy sectors}, electric power systems and heating systems. Both require real-time power balance, with the temperature dynamics in heating systems having much higher inertia arising from their large heat capacity. Thus, if controlled properly, heating systems can provide important support (like battery energy storage) to the frequency control mechanisms in power systems, with low costs and small user disruption.

However, as heat pump capacity increases, their impact on frequency dynamics becomes nonnegligible.
First, ensuring the stability of combined heat and power systems is important in the presence of renewables and load fluctuations. This challenge becomes even more significant due to the coupling of the diverse dynamics between the two interconnected energy systems. Moreover, achieving an appropriate power distribution in real time among the different electric and heat sources
contributing to frequency regulation is also important.   Inappropriate power sharing can cause operational problems and reduce the willingness to provide frequency regulation services.

\subsection{Literature survey}

{Recent studies demonstrate that modern variable-speed heat pumps have fast response capabilities, with sub-second responsiveness to control signals~\cite{ref_mit1,ref_mit2} and ramping to full output within only a few seconds~\cite{r1_inte}. By contrast, in local district heating systems, it may take from tens of seconds to minutes for power adjustments of heat pumps to translate into temperature changes and reach steady state. This motivates exploiting the thermal inertia and large capacity of heating systems to provide frequency support and facilitate renewable integration~\cite{r1_enable, r1_mlearn}: When heat pumps contribute to frequency regulation, other heat sources can adjust to compensate for their heat output, enabling water-based heating systems to mitigate frequency fluctuations from renewable generation, similar to battery storage.
In this way, low-cost and reliable renewable integration can be achieved. However, the fast response of heat pumps can also introduce fluctuations into the heating system, raising stability concerns and economic challenges related to supply-demand balance and power sharing among heat sources. Addressing these issues requires a control framework for heat pumps and the two interconnected energy sectors, explicitly accounting for both electrical and thermal dynamics.}

{
A parallel line of work explores the flexibility of heating system in economic dispatch.
For example, papers \cite{li2015combined} and \cite{li2015transmission} pioneered the concept of combined heat and power dispatch, uncovering substantial potential for renewable energy integration, followed and implemented in both academia and industry. In contrast to those setpoint-optimization frameworks, our work addresses frequency control at a faster time scale, focusing on small adjustments around the setpoints determined by economic dispatch.}

A number of studies have explored heat pumps contributing to frequency regulation under the assumption that their actions do not influence the power system dynamics. {For example,} papers \cite{ref_mit2, ref_mit1, ref_spnon1, ref_spnon2} and {\cite{r1_inte}} have focused on modeling individual heat pumps using experimental data, exploring {the inner elements with a heat pump},  and developing control strategies for frequency regulation.

Various studies~\cite{ref_mpn1, ref_mpn2, ref_mpn3, ref_mpn4} have explored how multiple heat pumps can provide frequency regulation services to the grid, with different control schemes having been explored
including on-off control~\cite{ref_onoff1, ref_onoff2} and Proportional-Integral-Derivative control~\cite{ref_pid1, ref_pid2}.
However, these studies do not provide a stability analysis in a general power network and in the more involved case where there is coupling with heating networks.

To address the challenges associated with the stability analysis, {some works adopt a single equivalent block model to consider frequency dynamics \cite{ref_spsn1,ref_spsn2,r1_multi,r1_adaptive}, or estimate frequency dynamics via data-driven approaches \cite{ref_other1_mpc}.
Although those approaches contribute to a stability insights}, they fail to accurately represent the dynamics of power networks comprising a large number of buses and transmission lines. At the same time, {they overlook the fluctuations caused to the interconnection to a heating network.}

Recent research~\cite{ref_machado} has modeled the dynamics and shown passivity properties of heating networks in a timescale of seconds, illustrating their potential in frequency regulation. Based on the heating network model in~\cite{ref_machado}, \cite{pump_schiffer} introduces models for the dynamics of combined heat and power networks. However, the challenge of achieving power sharing in both energy sectors in real time, {while also requiring a prior knowledge of steady-state equilibria and load disturbance}, remains unresolved.
Additionally, the aforementioned studies focus on specific generation dynamics, but practical applications may involve diverse generation dynamics of higher order, which should be taken into account in the analysis.

\subsection{Contributions}
In this paper, we consider heat pumps that are part of a district heating system and provide an ancillary service to the power grid by contributing to primary frequency regulation. Our contributions can be summarized as follows.

We first introduce a novel scheme for optimal power sharing among the different heat sources in the heating network. This is based on the use of the average temperature as a control signal and achieves optimal power sharing without {requiring the measurement of load disturbance or the prior knowledge of equilibrium}. We also present two schemes for heating systems to contribute to frequency regulation via heat pumps in power networks, {in which the heat pump can act as frequency-dependent load and converter-linked load}.
{We will show the stability analysis of the combined heat and power network under the two proposed control schemes, respectively, with a general network topology is considered. Furthermore, we will present how optimal power sharing can be achieved across different electric and heat sources.}
We then incorporate various classes of generation dynamics within our framework with stability and optimal power sharing guaranteed. Finally, we demonstrate via case studies the effectiveness of the proposed method under practical operational conditions with advanced system  dynamics, and various tradeoffs in the transient response are also discussed.

The paper is structured as follows. In section \ref{sec.prelim} we provide various preliminaries {to introduce problem setting}. In section \ref{sec.model} we describe the combined heat and power network that will be considered. In sections \ref{sec.stability} and \ref{sec.sharing} we present the main results associated with stability and power sharing.
Extensions associated with general passive power and heat generation dynamics are presented in section~\ref{sec.generalpas}. Finally, conclusions are drawn in section \ref{sec.conclusions}. {The {detailed} proofs can be found in Appendix~\ref{sec.appendix}}

\section{Preliminaries}\label{sec.prelim}
\begin{figure}[H]%
    \centering
    \includegraphics[trim = 0mm 0mm 0mm 0mm, clip, width = 0.9\columnwidth]{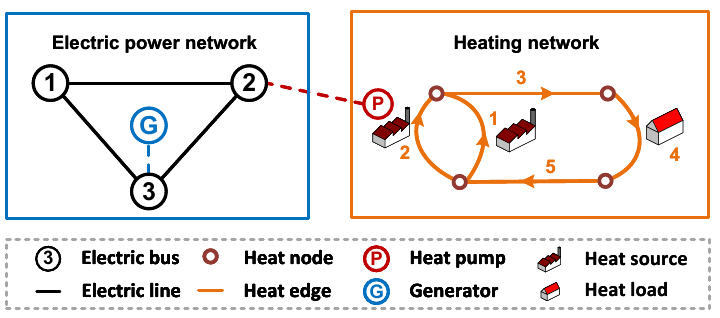}
      \caption{{Network topology graph of an example combined heat and power test system. The blue and red dotted lines, representing the connections between buses and generators/heat pumps, are not part of the main topology.}}
    \label{fig:3bus}
\end{figure}

{In this work, we consider heat {pumps providing} primary frequency regulation service to {a power grid}.
We consider a 
combined heat and power system, {where district heating systems connect to the power grid at sub-transmission or transmission level.} 
Each heating system connects to the power system via a bus, through either a large heat pump or aggregated co-located heat pumps, to provide ancillary services, as exampled in Figure~\ref{fig:3bus}.} For clarity, we omit the indexing of multiple heating networks in the modeling sections and demonstrate the framework's ability to accommodate them in Section~\ref{sec.simu} via case studies.

{As shown in Figure~\ref{fig:3bus}, as a heat source in the heating network, when the heat pump {contributes to} 
 frequency control 
will also produce fluctuations to the heating {system.} 
 To model {the dynamics of the combined heat and power network} 
 we consider the
 frequency dynamics of the power system and the thermal dynamics of the heating system, which will be introduced in Sections~\ref{sec.model}. We will propose {a control framework} for electric sources, heat sources, and heat pumps to tackle these {fluctuations.}}

{
{The interconnections in 
the combined heat and power system are described by means of directed graphs.} 
For the electric power network, we define a directed graph $(N_e, E_e)$, where $N_e = \{1, \dots, n_{N_e}\}$ is the set of buses and $E_e \subseteq N_e \times N_e$ is the set of transmission lines. Each link $(i,j) \in E_e$ represents a line between buses $i$ and $j$. We use $i:~i \rightarrow j$ and $k:~j \rightarrow k$ to indicate the predecessors and successors of bus $j$, respectively. The subset $N_e^G \subseteq N_e$ denotes generator buses, and $H_e \subseteq N_e$ denotes buses to which heat pumps are connected.}

{
For the heating network, we define a directed graph $G_h = (N_h, E_h)$, where $N_h$ is the set of heat nodes and $E_h = {1, \dots, n_{E_h}}$ is the set of heat edges. Each edge represents a single-input-single-output device such as a pipeline and a heat source. Specifically, we denote edges associated with heat pumps by $H_h$, with conventional heat sources (e.g., gas boilers) by $E_h^G$, and with heat loads by $E_h^L$. We use the temperature vectors $T^E = col(T^E_{i \in H_h}, T^E_{i \in E_h^G}, T^E_{i \in E_h^L})$ to collect the corresponding temperatures. Node temperatures are denoted by vector $T^N$, and the aggregated system temperature vector is $T = col(T^E, T^N)$.
}

{
{We adopt standard assumptions commonly used in the literature to analyze frequency dynamics:} 
bus voltage magnitudes in the electric network are approximately $1$ per unit, transmission lines $(i,j) \in E_e$ are lossless with susceptances $B_{ij} > 0$, and reactive power does not affect bus voltage angles or frequencies. The definitions {basic of mathematical} notations
have been moved to 
Appendix~\ref{appendix.mathdef}
 {to facilitate the readability}.}

\section{Model Formulation}\label{sec.model}
{
In this section, we present the model of the combined heat and power network. The dynamics of the power system and the heating system are introduced in Subsection~\ref{subsec.edynamics} {and 
in Subsection~\ref{subsec.hdynamics},} respectively. To address fluctuations arising from renewable generations, electric and heating loads, and heat pumps, we design control mechanisms for both electric and heat generators in Subsection~\ref{subsec.gencontrol}. In Subsection~\ref{subsec.pumpm}, we further propose two schemes through which heat pumps contribute to frequency regulation services. Building on these formulations, the subsequent Sections~\ref{sec.stability} and~\ref{sec.sharing} present 
{results} on system stability and power sharing, respectively.}

\subsection{Electric power system} \label{subsec.edynamics}
\begin{subequations} \label{eq.edynamics}
We will be using the swing equation{~\eqref{eq.wdynamics}~\eqref{eq.line1}} to describe the frequency dynamics at each bus $i$ in the electrical power network.
\begin{align}
 M_j \dot \omega_j &=  -p^L_j - p^P_j + p^G_j  - p^U_j  + \sum_{i:~i \rightarrow j} p_{ij} - \sum_{k:~j \rightarrow k} p_{jk}, \label{eq.wdynamics} \nonumber \\
 &~~~~~ \quad \qquad \quad \qquad \qquad  \qquad j \in N_e  \\
 \dot \eta_{ij} &= \omega_i - \omega_j, ~~~~\qquad \qquad \quad ~(i,j) \in E_e \label{eq.line1}\\
p_{ij} &= {B_{ij} \sin(\eta_{ij})}, \qquad \qquad ~~~(i,j) \in E_e. \label{eq.line2}
\end{align}
\end{subequations}
where variable $p^U_j=D_j\omega_j$ represents frequency dependent uncontrollable loads at bus $j$. $p^P_j = 0$ if there is no heat pump connected to bus $j$.
Variables $p^G_j$, $p^U_j$, $p^P_j$, $p^L_j$ denote deviations from nominal values with $p^G$, $p^U$, $p^P$, $p^L$, denoting corresponding vectors, respectively.
If bus $j$ is a load bus, $M_j = 0$, and $p^G_j = 0$. {The frequency {$\omega_j$ at different buses 
will converge to a synchronous value at steady-state, although this can be different at each bus transiently.}}

\subsection{Heating system} \label{subsec.hdynamics}
In this section, we describe the thermal dynamics of the heating system based on the model in~\cite{ref_machado}. {As shown in Figure~\ref{fig:3bus}, the dynamics of {the} heating network includes edge dynamics and node dynamics.
The edge dynamics, given in equation~\eqref{eq.edge_s}, describe the temperature dynamics {at} a heat source, load, or pipeline. Specifically,} the rate of change of the edge temperature is dependent on the temperature difference between its inlet and outlet as well as the heat power consumed/generated at {this edge.}

{The node dynamics, expressed in equation~\eqref{eq.edge_s}, describe the dynamics of a heat node when heat edges inject mass flow into it, as in the case of edges 1 and 2 {connected to} 
a common node in Figure~\ref{fig:3bus}. It indicates} that the rate of change of the thermal energy at node $k$ (left hand side) equals its net sum thermal energy input flows (right hand side). {Together, the dynamic behaviours of edges and nodes {and their corresponding heat capacity} provide the heating network with the flexibility to accommodate fluctuations from renewable generation in the power grid.}
\begin{subequations}
\begin{align}
    \rho C_{p} V^E_j \dot{T}^E_{j}&=\rho C_{p} q^E_j \left({T}^N_k-{T}^E_{j}\right) + h^G_{j} + h^P_{j}- h^L_{j}, \nonumber \\
            & \qquad \qquad \qquad j \in E_h, \  k \in N_h \label{eq.edge_s} \\
    \rho C_{p} V^N_k \dot{T}^N_k &= \sum_{j \in \mathfrak{T}_{k}} \rho C_{p} q^E_j T^E_j-(\sum_{j \in \mathfrak{T}_{k}}  \rho C_{p} q^E_j ) {T}^N_k,  \nonumber \\
            & \qquad \qquad \qquad  k \in {N_h}, \label{eq.node_s}
\end{align}
\end{subequations}
{{For simplicity  of presentation we consider  variables $h^P_j,  h^L_j, h^G_j$ at each edge $j$, and}}
if edge $j$ is associated with a heat pump, $h^G_j = h^L_j = 0$;
Similarly, if  edge $j$ is a non-pump heat source, $h^P_j = h^L_j = 0$, and if edge $j$ is a heat load, $h^G_j = h^P_j = 0$ whereas $h^L_j$ is a given constant. $C_{p}$ and $\rho$ are the {heat} capacity and density of water, respectively. In the normalized per-unit system, $\rho C_{p} = 1$. Set $\mathfrak{T}_{k}$ indicates the edges that inject flow to node $k$.

 For convenience in the analysis, we write equations~\eqref{eq.edge_s} and~\eqref{eq.node_s} into matrix form:
\begin{align}\label{eq.hdynamics}
V \left[\begin{array}{c}\dot{T}^{{E}} \\ \dot{T}^{N}\end{array}\right]
= - A_h\left[\begin{array}{c}T^{{E}} \\ T^{N}\end{array}\right]+  \left[\begin{array}{c} h^G+h^P-h^L \\ \textbf{0} \end{array}\right] ,
\end{align}
where $h^P$, $h^G$, and $h^L$ are $n_{E_h}$ dimension vectors with elements $h^P_j$ , $h^G_j$, and $h^L_j$, respectively.
$V = diag(\{V^E_j\}, \{V^N_j\})$ indicates the volume matrix.

Note that
\begin{align} \label{eq.A}
   { A_h=\left[
\begin{array}{cc}diag(q^E) & -diag(q^E) B_{sh}^{T} \\
-B_{th} diag(q^E)              & diag(B_{th} q^E)\end{array}\right]}
\end{align}
is a constant matrix for a given mass flow vector $q^E$, where
$q^E = col(q^E_j)$. $B_{th} = \frac{1}{2} \left(|B_{h}| + B_{h}\right)$, and $B_{sh} = \frac{1}{2} \left(|B_{h}| - B_{h}\right)$. {$B_h$ is the incidence matrix of the heating system, where $B_{h, ij} = 1$, if the flow through edge $j \in E_h$ injects into node $i \in N_h$; $B_{h, ij} = -1$, if the flow through edge $j \in E_h$ originates from node $i \in N_h$; $B_{h, ij} = 0$, otherwise}. It can be shown that $A_h+A^T_h$ is positive semidefinite with a simple eigenvalue at the origin and $A_h$ satisfies $A_h\textbf{1}=0, \textbf{1}^TA_h=0$ \cite{ref_machado}.

 \subsection{Generation control} \label{subsec.gencontrol}
Generators contribute to frequency control in the power grid. We consider heat pumps that operate in a mode that provides an ancillary service to the power grid by contributing to frequency control. The heat pumps are also part of a heating network. In particular, we have the following frequency regulation schemes for generators:
\begin{subequations}\label{eq.control}
\begin{align}
   {\tau_{e,j} \dot p^G_j} &= -p^G_j - \tilde Q_{e,jj} \omega_j, \quad j \in N_e^G \label{eq.controle}
\end{align}
where {$\tau_{e,j}$ is the 
{time constant} of generator at bus $j$.}
$\tilde Q_{e,jj} = \frac{1}{Q_{e,jj}}$.
For convenience, we define two diagonal matrices $Q_{e} := diag(\{Q_{e,jj}\})$ and $\tilde Q_{e} := diag(\{ \tilde Q_{e,jj} \})$.

We consider also conventional heat sources in the heating system contributing to temperature control via the scheme:
\begin{align}
    {\tau_{h,j} \dot h^G_j} &= -h^G_j - \tilde Q_{h,jj} \bar T , \quad j \in E_{h}^G \label{eq.controlh}
\end{align}
\end{subequations}
where {$\tau_{h,j}$ is the heat time constant of the source at edge $j$.}
$\tilde Q_{h,jj} = \frac{1}{Q_{h,jj}}$.
We also denote $Q_h=diag(\{ Q_{h,jj} \})$ and $\tilde Q_{h} := diag(\{ \tilde Q_{h,jj} \})$.
Variable $$ {\bar T} :=   \frac{\textbf{1}^T VT}{\mathcal{V}}$$
is the average temperature of the heating system, which is calculated as a weighted mean of temperature deviations.  The weights used reflect the energy stored in the nodes and edges of the heating system, where $\mathcal{V} = \textbf{1}^T V \textbf{1}$ is a constant equal to the sum of the heating system volumes.

\begin{rem}{(Average temperature)}
    {The significance of using {the} average temperature $\bar T$ in the control policies, rather than the local temperature $T$ or virtual variables, is inspired by {the following 
    considerations:
    It converges to a synchronous value at steady state (analogous to frequency) thus allowing power sharing to be achieved without a prior knowledge of demand; it allows to design control policies with stability guarantees for the combined heat and power network in general network topologies.}}
        {Stability and power sharing results based on the usage of the proposed $\bar T$ {as a control input} are presented in Sections~\ref{sec.stability} and \ref{sec.sharing}, respectively. {A comparison with more conventional 
        approaches where the local-temperature is used a control input are presented  in Section~\ref{subsec.case1}}.}
\end{rem}

\begin{rem}{(General dynamics)}
    First-order dynamics in~\eqref{eq.control} are used here for simplicity in the presentation. {For instance, similar generation control schemes for heat sources can be found in the literature~\cite{pump_schiffer}, and can represent the control of gas boilers as in~\cite{r1_boiler}. However, in practice, we acknowledge that the dynamics of electric and heat generation units can vary under different applications and are often more complex than first-order models. To address this,} in Section~\ref{sec.generalpas} we will generalize the first-order dynamics in~\eqref{eq.control} to general classes of input strictly passive systems from $\omega_j$ to $p^G_j$ and from $\bar T$ to $h^G_j$.
\end{rem}

\subsection{Heat pump} \label{subsec.pumpm}
{We introduce two proposed schemes through which heat pumps support {frequency regulation in the power network}. In both schemes, when heat pumps adjust their output to support frequency, the resulting heat imbalance is compensated by other {heat sources 
via adjustments as} described in equation~\eqref{eq.controlh}.}
{In section \ref{sec.sharing} we will show how using the average temperature as a control input allows to achieve an optimal power sharing.}

{
{We consider} that {the}} heat pump operates around an operating point, where the coefficient of performance (CoP), $C_{o,j},$ is approximately a constant~\cite{pump_kim}. The heat pump's electric power consumption and heat power generation are
\begin{align}
    h^P_{k_j} &=  C_{o,j} p^P_{j}, \quad j \in H_e,~k_j \in H_h \label{eq.pumpmodel}
\end{align}
where $j$ is the bus in the electrical network with a heat pump, and $k_j$ is the corresponding edge in the heating network it is associated with.

We consider two schemes for the heat pump participating in frequency regulation.
\begin{itemize}
    \item \textbf{Mode 1:} Frequency dependent load.
    In this mode, the heat pump works as a frequency dependent load as shown in~\eqref{eq.pumpm1} to provide frequency support for the power network.
    \begin{align}
        p^P_j &=  a_{1,j}\omega_j, \quad j \in H_e \label{eq.pumpm1}
    \end{align}
    where $a_{1,j}$ is a positive constant. Equation~\eqref{eq.pumpm1} indicates that we regulate the shaft speed of the heat pump in proportion to the bus frequency.

    \item \textbf{Mode 2:} Converter-linked load.
    In this mode, we consider the output of the heat pump converter as a separate bus with zero inertia, where the frequency is set by the heat pump.
    The scheme is given by equation~\eqref{eq.pumpm2}.
    \begin{subequations}\label{eq.pumpm2}
    \begin{align}
        0 &=  - p^P_j  + \sum_{i:~i \rightarrow j} p_{ij} - \sum_{k:~j \rightarrow k} p_{jk}, \quad j \in H_e  \label{eq.pumpm2_c2}\\
        \omega_{j} &=  m_j \bar T
        \label{eq.pumpm2_c1}
    \end{align}
    \end{subequations}
    {Coefficient $m_j$ links the frequency {of the heat pump bus in the electrical power network with the} 
    average temperature in Mode 2. {As it will be shown in section \ref{sec.sharing} it}
    influences the power sharing between the two energy sectors. In practice, this coefficient is determined through ancillary service market design, considering cost trade-offs between the two energy sectors and contractual arrangements.}
    Note that the heat pump power is dependent on the power transfer from other buses as shown in the power balance constraint~\eqref{eq.pumpm2_c2}.
\end{itemize}

\begin{rem}{{(Difference between the two modes)}}
    Mode 2 has two major differences from Mode~1. {First, shown in equation~\eqref{eq.pumpm2_c1}, the heat pump frequency is set directly by the converter, which serves as an intermediate variable to determine}
    the power consumed by the heat pump via the power transfer from neighboring buses (analogous to matching control schemes in hybrid AC/DC networks). This is in contrast to Mode 1 where the heat pump acts as a frequency-dependent load.
    Second, Mode 2 treats the heat pump converter as a separate bus, which implies that the underlying graph of the electric power system is different from that in Mode 1, {i.e., there is an additional bus associated with the frequency of the heat pump converter}.
\end{rem}

\section{Stability of Combined Heat and Power Network}\label{sec.stability}

In this section, we analyze the stability of the combined heat and power system under the two different participation modes of heat pumps described in the previous section, where the heat network also contributes to frequency control via the control policies of the heat pumps.

\begin{assum} \label{assum.equ}
The combined heat and power network \eqref{eq.edynamics}-\eqref{eq.pumpmodel}, together with~\eqref{eq.pumpm1} or~\eqref{eq.pumpm2}, respectively, admits an equilibrium
point $x^* = (\eta^*, \omega^*, p^{G\ast}, {T}^*, h^{G\ast})$ with $|\eta^*_{ij}| < \frac{\pi}{2}$,  $\forall ij \in E_e$.
\end{assum}
This assumption on $\eta^*_{ij}$ is a security constraint and is ubiquitous in the power network literature.

\begin{thm}[Stability] \label{thm:stability}
    Consider the combined heat and power network described by \eqref{eq.edynamics}-\eqref{eq.pumpmodel}, \eqref{eq.pumpm1} or~\eqref{eq.edynamics}-\eqref{eq.pumpmodel}, \eqref{eq.pumpm2}.
    Consider an equilibrium point of this system that satisfies Assumption~\ref{assum.equ}.
    Then, there exists an open neighborhood of this equilibrium point such that all solutions of the system starting in this region converge to an equilibrium point.
\end{thm}

The proof of Theorem \ref{thm:stability} is {based on} {Lyapunov arguments and is}
provided in Appendix \ref{proof.stability_m1} and \ref{proof.stability_m2}{: In} Appendix \ref{proof.stability_m1} we prove the theorem when Mode 1 is used, and in Appendix \ref{proof.stability_m2} we prove the theorem when Mode 2 is used.

\section{Power Sharing in the Combined Heat and Power Network}\label{sec.sharing}
In this section, we show that the use of the average temperature in the control policies in the heating system allows to achieve optimal power sharing. In Mode 1, we show that the power sharing is optimal in the electric and the heating system, separately. In Mode 2, we show that {optimal} {power sharing in the combined electric and heat network} 
is achieved among all electric and heat sources.

\vspace{-10pt}
\subsection{Mode 1}

\begin{prop}{(Optimality under Mode 1)}~\label{thm.opti_sep}
Consider the combined heat and power network described by \eqref{eq.edynamics}-\eqref{eq.pumpmodel} and~\eqref{eq.pumpm1}.
The deviation in the power generation $p^G$ and of the heat pump load $p^P$ at equilibrium after a step change in electrical load $p^L$ is the solution to the optimization problem
\vspace{-0.4em}
\begin{subequations}~\label{eq.opti_sepe}
\begin{gather}
    \min_{p^G, p^P, p^U} \frac{1}{2} (p^G)^T Q_e p^G
        +\frac{1}{2} (p^P)^T Q_p (p^P)^T + \frac{1}{2} (p^U)^T Q_u p^U
\end{gather}
\vspace{-1em}
\begin{gather}
    \text{s.t.} \quad \textbf{1}^T p^G = \textbf{1}^T p^L + \textbf{1}^T p^P + \textbf{1}^T p^U,  \label{eq.opti_ec1}
\end{gather}
\end{subequations}
where
$Q_p = diag(\{ Q_{p,jj} \})$ with $Q_{p,jj} = \frac{1}{a_{1,j}}$, $j\in H_e$
and $Q_u = diag(\{Q_{u,jj}\})$ with $Q_{u,jj} = \frac{1}{D_{j}}$, $j\in N_e$.\\

Given the power deviation of heat pump $h^P$ in the solution of \eqref{eq.opti_sepe}, the deviation in power $h^G$ of the conventional heat sources is the solution to the optimization problem
\begin{subequations}\label{eq.opti_seph}
\begin{align}
    \min_{h^G}  \frac{1}{2} (h^G)^T {Q}_h h^G, \quad  \\
    \text{s.t.} \quad \textbf{1}^T h^G = \textbf{1}^T (h^L - h^P).  \label{eq.opti_hc1}
\end{align}
\end{subequations}
\end{prop}
The proof of Proposition~\ref{thm.opti_sep} is provided in Appendix~\ref{proof.opti_sep}.

\vspace{-5pt}
\subsection{Mode 2}
\begin{thm}{(Optimality under Mode 2)}~\label{thm.opti_joint}
Consider the combined heat and power network described by \eqref{eq.edynamics}-\eqref{eq.pumpmodel} and~\eqref{eq.pumpm2}. Then the power contribution of the electric generators $p^G$ and conventional heat sources $h^G$ is the solution to the combined heat and power optimization problem
\begin{subequations}~\label{eq.opti_joint}
\begin{gather}
    \min_{p^G, h^G, p^U}\frac{1}{2} (p^G)^T Q_e p^G + \frac{1}{2} (h^G)^T \alpha_j Q_h h^G
    +  \frac{1}{2} (p^U)^T Q_u p^U, \label{eq.opti_jointobj}
\end{gather}
\vspace{-1.5em}
\begin{align}
    \text{s.t.} \quad \textbf{1}^T p^G &= \textbf{1}^T p^L + \textbf{1}^T p^P + \textbf{1}^Tp^U ,\label{eq.opti_jointc1}\\
    \textbf{1}^T h^G &= \textbf{1}^T h^L -  \textbf{1}^T h^P,   \label{eq.opti_jointc2} \\
 h^P_{k_j} &=   {C_{o,j} p^P_{j}, \quad j \in H_e}. \label{eq.opti_jointc3}
\end{align}
where coefficient $\alpha_j = \frac{m_j}{C_{o,j}},~j \in H_e$.
\end{subequations}
\end{thm}

The proof of Theorem~\ref{thm.opti_joint} is provided in Appendix \ref{proof.opti_joint}. {Theorem~\ref{thm.opti_joint} shows that by interlinking the thermal temperature with the pump converter frequency of the corresponding separate (virtual) bus, we can achieve appropriate power sharing between the combined heat and power network.} It should be noted that when multiple heating areas are considered, the second term in the objective function~\eqref{eq.opti_jointobj} will become the summation of the heat generation costs of all heating networks.

\section{General passive dynamics} \label{sec.generalpas}
In practical combined heat and power networks, the dynamics of generators and conventional heat sources are diverse, extending beyond first-order dynamics as in \eqref{eq.control}. To address this practical issue, we describe in this section how the approach in the paper can accommodate general higher-order generation dynamics characterized by input strictly passive systems, relating $\omega_j$ to $p^G_j$ and  $\bar T$ to $h^G_j$, respectively.

More precisely, in the power network we consider $p^G_j$ now as the output of a nonlinear system with input $\omega_j$, namely
\begin{subequations}\label{eq.epas}
    \begin{align}
    \dot x^s_{e,j} &= -f_{e,j}(x^s_{e,j},~\omega_j), \quad j \in N_e^G,   \\
             p^G_j &= -g_{e,j}(x^s_{e,j},~\omega_j), \quad j \in N_e^G.
\end{align}
\end{subequations}
where $x^s_{e,j}$ is the state of the system, and $f_{e,j}$ and $g_{e,j}$ are locally Lipschitz functions. System~\eqref{eq.epas} is said to be locally input strictly passive around the equilibrium point  $\omega^*$, $x^{s*}_{e,j}$, if there exist open neighborhoods $\Omega_{j}$ of $\omega^*$ and $X^s_{e,j}$ of $x^{s*}_{e,j}$ and a continuously differentiable positive semi-definite function $V^s_{e,j}(x^s_{e,j})$, such that for all $\omega_j \in \Omega_{j}$ and all $x^s_{e,j} \in X^s_{e,j}$,~$j \in N^G_e$, one has
\vspace{-0.3em}
\begin{align}
    \dot V_{e,j}^g(x^s_{e,j}) \leq (-\omega_j + \omega^* )(p^G_j - p^{G*}_j) - \phi_{e,j}(-\omega_j + \omega^*), \nonumber
\end{align}
where $\phi_{e,j}$ is a positive definite function. At the equilibrium point, we define
\vspace{-.2cm}
{\begin{gather}\label{eq:static_map}
p^{G*}_j = K_{p^G_j}(-\omega^*),
\end{gather}}
where $K_{p^G_j}(-\omega^*)$ is a static input-output characteristic map, which is defined under the assumption that the equilibrium value $p^{G\ast}_j$ is unique for a given constant input $\omega^\ast$. We also assume that the equilibrium point $x^{s*}_{e,j}$ is asymptotically stable for subsystem \eqref{eq.epas} for constant input $\omega^*$.

Similarly, we consider a general class of generation dynamics of conventional heat sources, where we define $h^G_j$ is the output of a nonlinear system
\begin{subequations}\label{eq.hpas}
    \begin{align}
    \dot x^s_{h,j} &= -f_{h,j}(x^s_{h,j},~\bar T), \quad j \in E^G_h,  \\
             h^G_j &= -g_{h,j}(x^s_{h,j},~\bar T), \quad j \in E^G_h.
\end{align}
\end{subequations}
where $x^s_{h,j}$ is the state of the system, and $f_{h,j}$ and $g_{h,j}$ are locally Lipschitz. System~\eqref{eq.hpas} is said to be locally input strictly passive around the equilibrium point $\bar T^*$, $x^{s*}_{h,j}$, if there exist open neighborhoods c{$\mathcal{T}_{j}$ of} $\bar T^*$ and  $X^s_{h,j}$ of $x^{s*}_{h,j}$ and a continuously differentiable positive semi-definite function $V^s_{h,j}(x^s_{h,j})$, such that for all c{$\bar T \in \mathcal{T}_{j}$} and all $x^s_{h,j} \in X^s_{h,j}$,~$j \in E^G_h$, one has
\begin{align}
    \dot V_{h,j}^g(x^s_{h,j})& \leq (-\bar T + \bar T^* )(h^G_j - h^{G*}_j) - \phi_{h,j}(-\bar T + \bar T^*),
\end{align}
where $\phi_{h,j}$ is a positive definite function. At the equilibrium point, we similarly define
{
\begin{gather}\label{eq:static_T}
h^{G*}_j = K_{h^G_j}(-\bar T^*),
\end{gather}
}
where $K_{h^G_j}(-\bar T^*)$ is a static input-output characteristic map and assume  $x^{s*}_{h,j}$ is asymptotically stable for subsystem \eqref{eq.hpas} {with constant input $\bar T^\ast$}.

\begin{assum}~\label{assum.equ_pa}
   Each of the systems defined in~\eqref{eq.epas} and \eqref{eq.hpas} with given inputs and outputs respectively are locally input strictly passive about 
   equilibrium values associated with an equilibrium point of the combined heat and power network. The corresponding storage functions also have strict local minima at this equilibrium point.
   \end{assum}

\vspace{-1.5em}
\subsection{Stability of Combined Heat and Power Network under General Passivity}
\begin{cor}{(Stability)} \label{thm.general_pas}
    Consider the combined heat and power network described by \eqref{eq.edynamics}-\eqref{eq.hdynamics},\eqref{eq.pumpmodel},\eqref{eq.pumpm1},\eqref{eq.epas},\eqref{eq.hpas}, or \eqref{eq.edynamics}-\eqref{eq.hdynamics},\eqref{eq.pumpmodel},\eqref{eq.pumpm2}, \eqref{eq.epas},\eqref{eq.hpas}. Consider an equilibrium point of this system that satisfies Assumptions~\ref{assum.equ} and~\ref{assum.equ_pa}.
    Then there exists an open neighborhood of this equilibrium point such that all solutions of the system starting in this region converge to an equilibrium point.
\end{cor}
\vspace{-0.5em}
A detailed proof of Proposition~\ref{thm.general_pas} can be found in Appendix~\ref{proof.general_pas}.

\vspace{-0.5em}
\subsection{Power Sharing under General Passive dynamics}
It can be shown that analogous power sharing results to those stated in Proposition \ref{thm.opti_sep} and Theorem \ref{thm.opti_joint} can be obtained when the generalized generation dynamics \eqref{eq.epas}, \eqref{eq.hpas} are considered. In particular, the power sharing results hold but with the quadratic terms $\frac{1}{2} (p^G)^T Q_e p^G$, $\frac{1}{2} (h^G)^T \frac{m_j}{C_{o,j}}Q_h h^G$ replaced by $\hat C_{e}(p^G)$, $\frac{m_j}{C_{o,j}}\hat C_{h}(h^G)$ respectively, where {functions $\hat C_e, \hat C_h$ are related to the equilibrium relations \eqref{eq:static_map}, \eqref{eq:static_T} as follows}
\begin{subequations}
    \begin{align}
        K_{p^G_j}(\omega^*) = (\hat C^{'}_{e,jj})^{-1}(\omega^*),~j \in N_e^G \\
        K_{h^G_j}(\bar T^*) = (\hat C^{'}_{h,jj})^{-1}(\bar T^*),~j \in E_h^G
    \end{align}
    \end{subequations}
and   $\hat C_{e}(p^G)$, $\hat C_{h}(h^G)$ are assumed to be strictly convex.
This follows from the Karush–Kuhn–Tucker (KKT) conditions for optimality as in the derivations {for the power sharing results} in Section \ref{sec.sharing}.

\section{Simulations - Case Studies} \label{sec.simu}
In this section, we present two case studies based on benchmark test systems. The first case study aims to demonstrate the effectiveness of the proposed approaches using the models analyzed in the paper.
The second case extends the proposed approaches to a more complex and practical scenario that incorporates more advanced power system dynamics, using the Power System Toolbox (PST)~\cite{ref_pst}.

\subsection{Case study 1: Example with frequency dynamics} \label{subsec.case1}
{In this case {study}, we compare the proposed two schemes (Mode 1 and Mode 2), and a comparison case (labeled ``CMP'') based on~\cite{pump_schiffer}. In the CMP case, as in~\cite{pump_schiffer}, heat pump power {is adjusted based on frequency deviations at} 
the connected bus, while heat sources use the local edge temperature $T^E$ rather than our proposed average temperature $\bar T$ as the control variable. To ensure a fair comparison, the power system generation in the CMP case is set the same as in our schemes of~\eqref{eq.controle}, making CMP equivalent to Mode~1 on the power system side. This allows us to focus on the role of the proposed control and power sharing schemes in the heating system. The comparison highlights the advantage of using the proposed average temperature {as a control input}.}

\begin{figure}[h]%
    \centering
    \includegraphics[trim = 0mm 0mm 0mm 0mm, clip, width = 0.9\columnwidth]{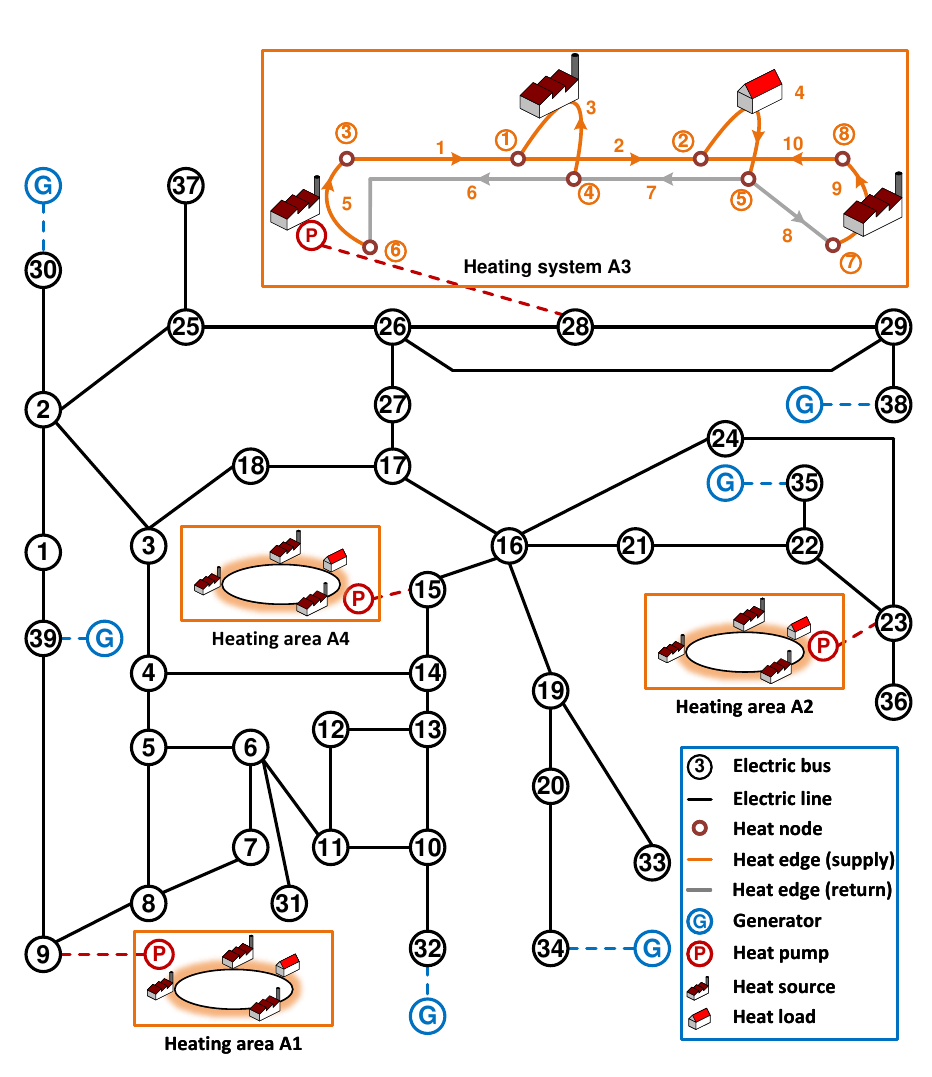}\label{fig.fmax}
      \caption{{Network topology graph of the combined heat and power test system with a 39-bus power system and 4 heating areas. For clarity, the internal topologies of heating areas 1, 2, and 4 are omitted.}}
    \label{fig:39bus}
\end{figure}

This simulation is performed using the modified IEEE-39 bus system~\cite{ref_ieee39og}, which is based on the ISO New England system. The generator inertia and bus damping coefficients are designed using data in~\cite{ref_ieee39r2}. As shown in Figure~\ref{fig:39bus}, the model includes four heating areas, each connected to a different bus via a heat pump that provides frequency support.

The power system base power is set to 100 MVA as in \cite{ref_ieee39og}, with a total load of around 65 GW. We consider each heating area at a neighborhood scale similar to those in West Cambridge~\cite{ref_camheat}, with each system having a base power of 100 MW and a total load of around 80 MW. In the simulation, a 0.5 p.u. generation loss is applied at buses 30 and 32, respectively, and we analyze frequency control performance at various buses and heating areas. We set the coefficients in the control policies of the heat pumps in Mode 1 and 2 such that both modes have the \textit{same frequency response} at steady-state.

\begin{figure}[h]%
    \centering
    \subfloat[~]{
		\includegraphics[trim = 1mm 0mm 1mm 3mm, clip, width = .8\columnwidth]{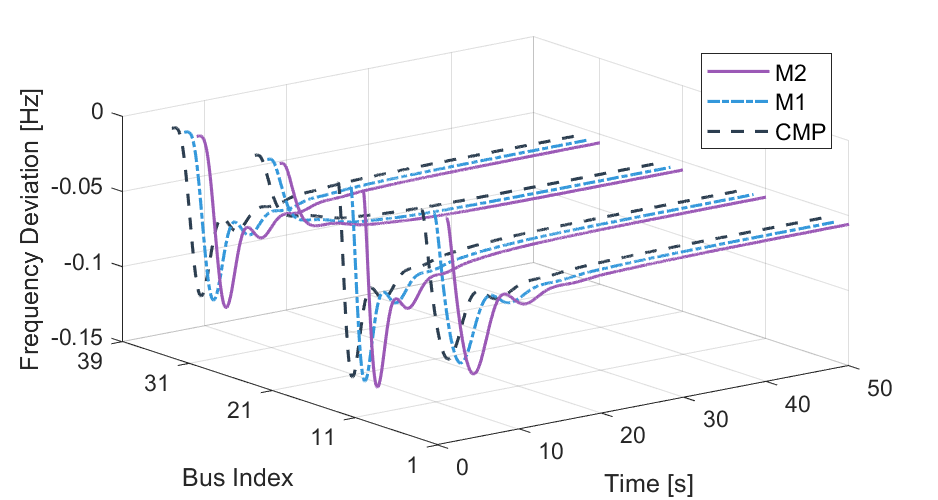}\label{fig:freq2}
	}\\
    \subfloat[~]{
		\includegraphics[trim = 1mm 0mm 4mm 5mm, clip, width = .49\columnwidth]{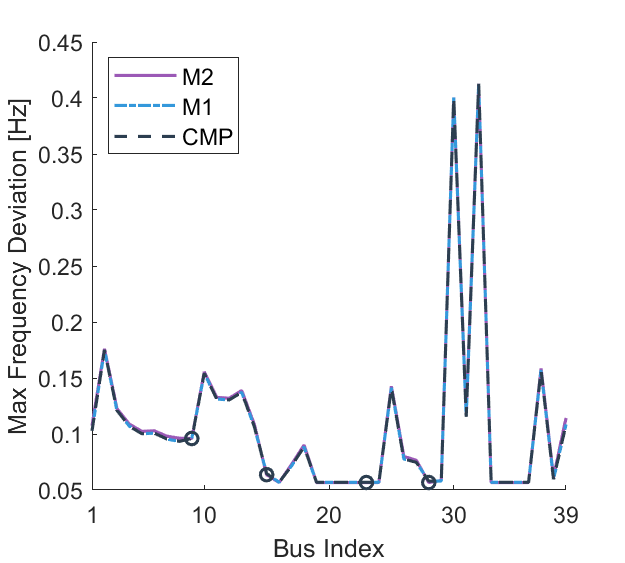}\label{fig:freqmax}
	}
    \subfloat[~]{
		\includegraphics[trim = 1mm 0mm 4mm 5mm, clip, width = .49\columnwidth]{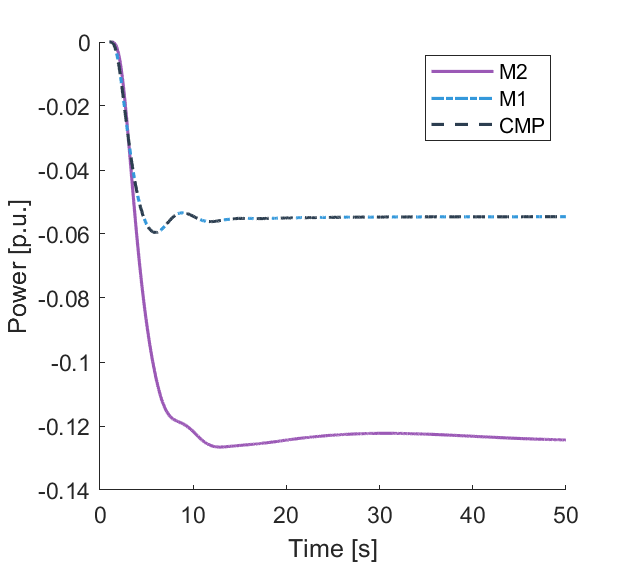}\label{fig:pupsum}%
	}
  \caption{(a) Frequency deviation {at buses 1, 11, 21, 31},  (b) Maximum frequency deviation throughout the trajectory. Circles indicate the buses connected to heat pumps, and (c) Total electric power adjustment by heat pumps across all heat areas.}
    \label{fig:fall}
\end{figure}

\begin{figure}[h]%
	\centering
    \subfloat[~]{
		\includegraphics[trim = 0mm 0mm 0mm 0mm, clip, width = .49\columnwidth]{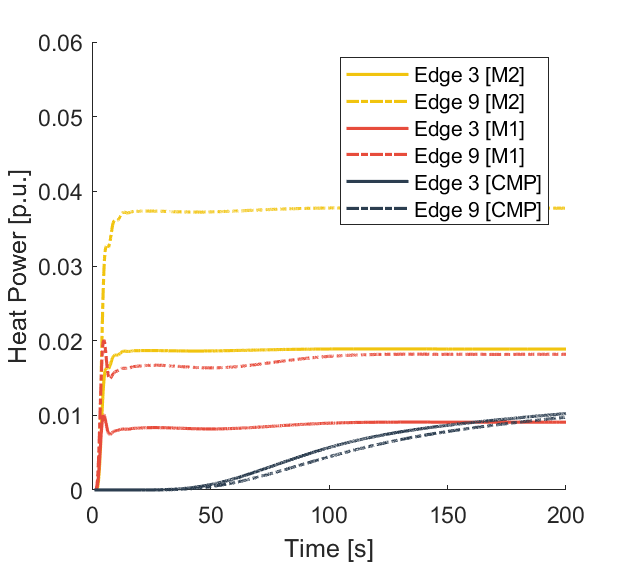}\label{fig:harea4}%
	}
     \subfloat[~]{
		\includegraphics[trim = 0mm 0mm 0mm 0mm, clip, width = .49\columnwidth]{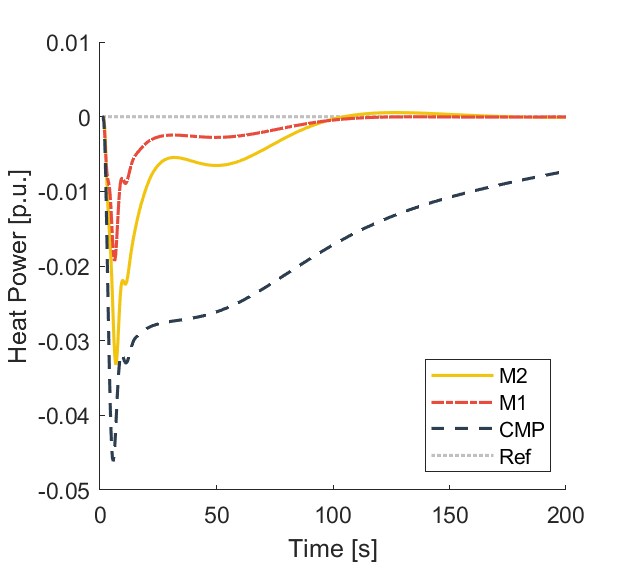}\label{fig:imb}%
	}
  \caption{{(a) Heat output from conventional heat sources in Area 4, {where the cost coefficients of edge 3 and 9 are 2 and 1, respectively}, and (b) Heat power imbalance between generation and demand in area 4. The legend entry REF denotes the zero imbalance reference curve.} }
    \label{fig:case_freqonly}
\end{figure}

The results in Figure~\ref{fig:freq2} demonstrate that both Mode 1 and Mode 2 {effectively stabilize the frequency. By comparing Figures~\ref{fig:fall} and~\ref{fig:case_freqonly},} the two modes exhibit asymmetric effects: Mode 1 is more responsive and can provide faster frequency support with smaller aggregate power contribution from the heating system; Mode 2 responds more slowly but can achieve system-wide optimal power sharing.

{Regarding the frequency response, as shown in Figure~\ref{fig:freq2}}, in Mode 1, heat pumps act as frequency dependent loads, enabling rapid frequency stabilization at bus 30 within 9.77 seconds, {while} Mode 2 takes a slightly longer time (11.9 seconds) to reach steady state.  {As {the} heat pump in the CMP case operates as frequency dependent load, its response is the same as Mode 1.} {Moreover, we found {that heat pump participation will} influence but not dominate frequency response as shown in Figure~\ref{fig:freq2}. This is because their contribution is limited to small adjustments around the setpoints, with each heat pump providing less than 0.035 p.u. (Figure\ref{fig:pupsum}), a value much smaller than that of conventional generators.} We also found that both modes have similar maximum frequency deviations over the trajectory as shown in Figure~\ref{fig:freqmax}. {Additionally, as shown in Figures~\ref{fig:freq2} and~\ref{fig:freqmax}, different buses have differences in frequency response, which demonstrates the importance of considering a power network model rather than simplifying {the aggregate power system to a single bus.}}

Meanwhile, {regarding the power sharing properties, as shown in Figure~\ref{fig:case_freqonly},} in Mode 1, the fast adjustment of heat pumps acts like a load fluctuation from the perspective of the heating system, causing disturbances in the heating system as illustrated in Figure~\ref{fig:harea4}.
In contrast, {heat pumps in Mode 2 {follow} the control scheme that} links converter frequency to the average temperature, 
{and imbalances in the heating system are recovered more quickly,
while} still providing effective frequency support. {Figure~\ref{fig:imb} shows {that 
Modes 1 and 2 eliminate} supply-demand imbalance at steady state.} Additionally, we see that both Mode 1 and 2 achieve power sharing in the heating system proportional to the cost coefficients, as illustrated in Figure~\ref{fig:harea4}. It should be noted that Mode 2 enables power sharing across the entire combined heat and power system, whereas Mode 1 achieves power sharing independently within each energy system as shown in Section~\ref{sec.sharing}.

{Compared to Modes 1 and 2, the CMP case shows slower response, inefficient power sharing, and supply-demand imbalance {for longer periods during transient operation}. This is because, although the heat pump power fluctuation is the same as in Mode 1, the CMP scheme relies solely on local temperature for heat source control, requiring disturbances to propagate through the network before being addressed. In contrast, our schemes use the average temperature, which captures real-time global information across the heating network and serves as a synchronous variable (analogous to frequency), thereby enabling a much faster response.}

\vspace{-0.5em}

\subsection{Case {study} 2: {Example with more advanced dynamics}}
Case study 1 in Section~\ref{subsec.case1} demonstrated the {proposed} stability and power sharing properties under the {considered dynamic models in Section~\ref{sec.model}}.
In a practical setting with various dynamics at different time scales, including those of voltage, generators, excitation systems, governors, transmission lines, and loads, might overlap and interfere with the system frequency response.

To address practical operational concerns and assess the effectiveness of the proposed methods in more realistic scenarios, we validate the proposed framework using the PST toolbox~\cite{ref_pst} that incorporates these complex dynamic models.

\begin{figure}[h]
    \centering
    \includegraphics[trim = 0mm 0mm 0mm 0mm, clip, width = 0.9\columnwidth]{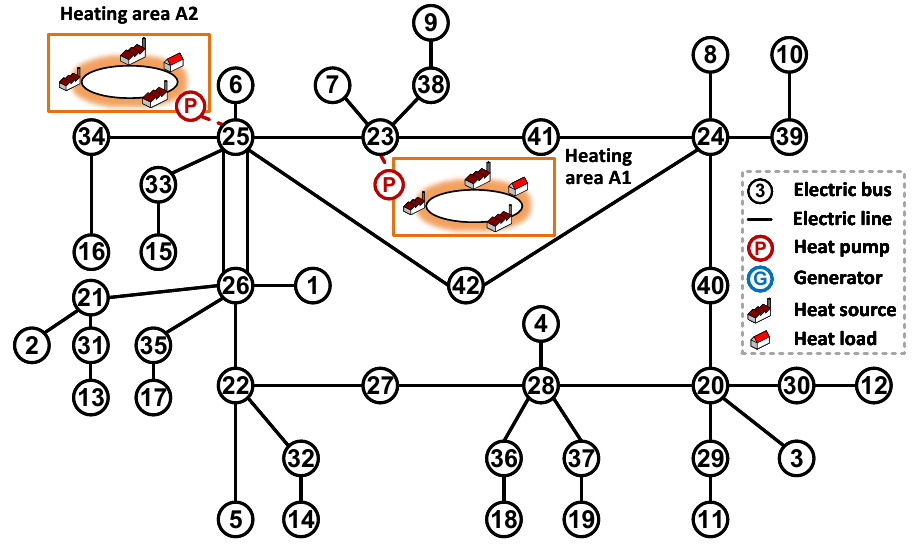}
      \caption{{Network topology graph of the combined heat and power test system with a 42-bus power system and 2 heating areas.}}
    \label{fig:42bus}
\end{figure}

{The simulations are conducted on the 19-machine benchmark case, \textit{d19k}, in PST~\cite{ref_pst}. We consider a combined heat and power system that integrates the 19-machine case as the power network together with two heating networks. As shown in Figure~\ref{fig:42bus}, the power system consists of 42 buses and 19 generators, with a total load demand of approximately 70 GW. The base power is set to 100 MVA. The network topology has a closed loop of interconnected buses representing the 230/500 kV buses, while the radial branches extending from it correspond to 20 kV buses. Two heating systems are integrated via aggregated, co-located heat pumps connected at transmission-level buses 23 and 25, respectively.} At 1 second, a step change is applied to the active power load at bus 23, reducing it by 5.68 p.u. from an initial value of 11.68 p.u., while the heating system loads remain unchanged.
We set the heat pump coefficients such that both modes have the \textit{same aggregate power adjustment} within the heating systems of 1.01 p.u. at steady-state. {Generators at buses 15 and 16 follow our generation control {scheme~\eqref{eq.controle}} for governor control, while the remaining generators operate under their original settings.}

\begin{figure}[H]%
	\centering
   	\subfloat[~]{
		\includegraphics[trim = 0mm 0mm 0mm 0mm, clip, width = .49\columnwidth]{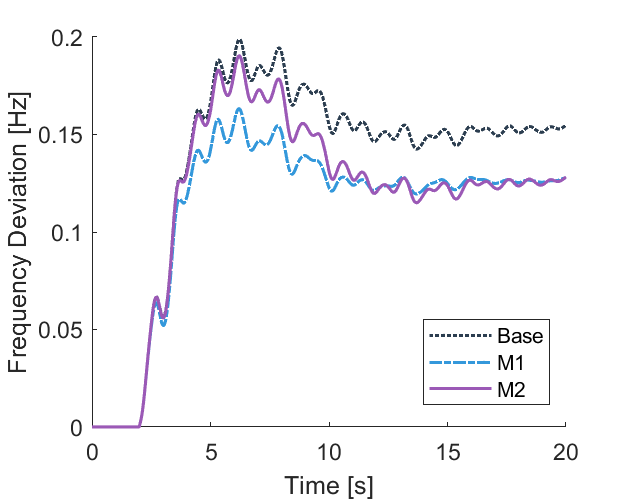}\label{fig:pst_f}}
 	\subfloat[~]{
		\includegraphics[trim = 0mm 0mm 0mm 0mm, clip, width = .49\columnwidth]{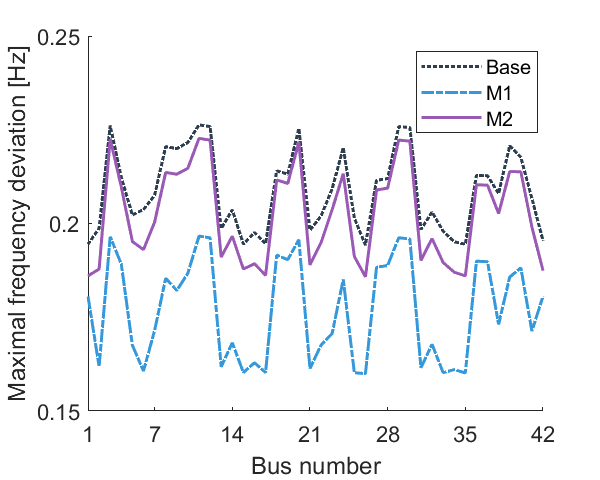}\label{fig:pst_maxf}
	}
    \caption{(a) Frequency deviation {at} bus 15. (b) Maximum frequency deviation over {the} trajectory. The base powers of the electric power system and the heating systems are 100 MVA and 100 MW, respectively. The legend `Base' indicates the base case, where the heat pumps do not provide any frequency support. }
    \label{fig:hall}
\end{figure}
\vspace{-0.6em}

As illustrated in Figure~\ref{fig:pst_f}, both Mode 1 and Mode 2 effectively stabilize the frequency to nearly identical values. Modes~1 and 2 continue to exhibit asymmetric effects as shown in Figures~\ref{fig:pst_f} and~\ref{fig:pst_maxf}: Similar to the results in Subsection \ref{subsec.case1}, Mode 1 is more responsive and can effectively reduce maximum frequency deviation, while Mode 2 provides a relatively slower frequency support but can achieve system-wide optimal power sharing.

{Unlike Subsection~\ref{subsec.case1}, in this case we observe 
{differences} in {the} maximum frequency deviation across all buses.
Mode 1 achieves a lower maximum frequency deviation due to its faster response {as a frequency-dependent load}. This difference can also be attributed to the lower damping coefficient of the 19-Machine system compared to the IEEE-39 bus system used in Case 1 in Subsection~\ref{subsec.case1}. Furthermore, the influence of additional system dynamics, such as transmission line and generator dynamics, highlights the importance of quickly responding to local frequency {deviations} 
which is a key advantage of Mode 1.} {Additionally, we found} both modes have nearly identical maximum {voltage deviations, with {1.68\% and 0.92\%} lower 
{compared} to the base case} throughout the {trajectories} across all buses, indicating {the} {effectiveness of the} proposed control schemes {under complex {power grid dynamics}}.

\section{Conclusion}\label{sec.conclusions}
In this paper, we have proposed two schemes, {as frequency-dependent load, and converter-linked load}, through which heat pumps can contribute to frequency control with the combined heat and power network maintaining its stability, while also achieving appropriate power sharing.
We have shown that the use of the average temperature as a control signal in a district heating system can lead to optimal power sharing without requiring a prior knowledge of the disturbances, while maintaining stability guarantees in general network topologies and general generation dynamics.

Simulation results show that the two proposed approaches, provide system operators with flexible options for frequency regulation. Heat pumps operating in {converter-linked} mode allow power sharing across all energy sources in the combined heat and power system. On the other hand, heat pumps in {frequency-dependent} mode achieve more responsive frequency stabilization, with power sharing achieved independently within each area.

{Future work will explore the extension of the proposed schemes to secondary frequency control, enabling coordinated multi-timescale regulation across energy networks. 
{At the same time,} further research on ancillary service market design and pricing mechanisms is essential to incentivize the adoption of heat pumps as reliable providers of primary and secondary frequency response.}

\section{Appendix} \label{sec.appendix}

\subsection{Mathematical Definitions} \label{appendix.mathdef}
We use $\textbf{0}_{n}$ and $\textbf{1}_{n}$ to denote the
$n$-dimensional vector of zeros and ones, respectively. For simplicity, in the presentation we will omit the subscript in the text.
For a vector $x\in \mathbb{R}^n$, we use $diag(x)$ to denote a diagonal matrix with elements $x_i$ in its main diagonal. For vectors $x\in\mathbb{R}^n$, $y\in\mathbb{R}^m$ we use the notation $col(x, y)$ to denote the vector $[x^T \ y^T]^T $. For a function $f(x)$, we use $f'(x) = \frac{df(x)}{dx}$ to denote its first-order derivative. We use $f^{-1}(y)$ to represent the preimage of the point $y$ under the function $f$, i.e., $f^{-1}(y) = \{x:~ f(x) = y\}$.

\subsection{Proof of Theorem \ref{thm:stability} under Mode 1} \label{proof.stability_m1}

We consider the combined heat and electrical power network described by \eqref{eq.edynamics}-\eqref{eq.pumpmodel}, \eqref{eq.pumpm1}.
We will first prove convergence to an equilibrium point for the states of the electrical power system, then we will prove convergence of the average temperature $\bar T$ in the heating system, and we will finally prove convergence of the temperature vector $T$ in the heating system.

The arguments for the convergence in the power system are analogous to those in \cite{kasis2016primary}, though to deduce convergence in the heating system requires more involved arguments due to the non-unique equilibrium points in the latter.\\

For the dynamics of the electric power system, the state vector is $x_e = (\eta, \omega,  p^{G}$). We consider storage functions $V_{1,\omega}(\omega) = \frac{1}{2} \sum_{j \in N_e} M_j (\omega_j - \omega^*)^2 $ and $V_{1,l}(\eta) = \sum_{ij \in E_e} B_{ij} \int_{\eta_{ij}^*}^{\eta_{ij}}(\sin \theta  - \sin \eta_{ij}^*) d\theta $  and then consider the time derivative of $V_{1,\omega}(\omega) + V_{1,l}(\eta)$ along the
system trajectories
\begin{align} \label{eq.V1wl}
&\dot V_{1,\omega}  
+ \dot V_{1,l}
= \sum_{j \in N_e} -D_j (\omega_j - \omega^*)^2   \\
&~~~~ +  \sum_{j \in N_e} (\omega_j - \omega^* )(p^G_j - p^{G*}_j)  - \sum_{{j \in H_e}} a_{1,j}(\omega_j - \omega^* )^2. \nonumber
\end{align}

We consider also the storage function $V_{1,g}(p^G) = \frac{1}{2} \sum_{j \in N_e^G}   \frac{1}{\tilde Q_{e,jj}}(p^G_j - p^{G*}_j)^2$.
Then
\begin{align}
&&\dot V_{1,g} =  \sum_{j \in N_e^G} - \frac{1}{\tilde Q_{e,jj}}(p^G_j - p^{G*}_j)^2 - (\omega_j - \omega^*)(p^G_j - p^{G*}_j). \nonumber
\end{align}

We define $V_{1,e}(x_e)= V_{1,\omega} + V_{1,l}+ V_{1,g}$ and hence have
\begin{align}
    & \dot V_{1,e}(x_e) =
    \dot V_{1,\omega} + \dot V_{1,l}+ \dot V_{1,g} \\
    &= \sum_{j \in N_e} -(D_j +a_{j}) (\omega_j - \omega^*)^2 - \sum_{j \in N_e^G} \frac{1}{\tilde Q_{e,jj}} (p^G_j - p^{G*}_j)^2,  \nonumber
\end{align}
where $a_{1,j} = 0$ for $j \notin {{H_e}}$.

We then consider $V_{1,e}$ as the candidate Lyapunov function for the electric power system
which from the analysis above is non-increasing with time.
From the condition that $\eta_{ij} < \frac{\pi}{2}$, the function $V_{1,e}(x_e)$ has a strict local minimum at $ x_e^* = (\eta^*, \omega^*, p^{G\ast})$. Hence we can choose a neighborhood $\mathcal{S}_1$ of $ x_e^*$ given by $\mathcal{S}_1=\{(\eta, \omega, p^{G}): V_{1,e} \leq \epsilon_1 \}$ for some sufficiently small $\epsilon_1 >0 $ such that this is a compact and positively-invariant set that contains $ x_e^*$.

We then apply LaSalle's Theorem, with $V_{1,e}$ as the Lyapunov-like function, which states that all trajectories of the system starting from within $\mathcal{S}_1$ converge to the largest invariant set within $\mathcal{S}_1$ that satisfies $\dot V_{1,e}(x_e) = 0$.

Note that coefficients $D_j$, $a_{j}$, $\tilde Q_{e,jj}$ are all positive, indicating $\dot V_{1,e}(x_e) = 0$  implies  $(\omega_j, p^G_j) = (\omega^*, p^{G*}_j)$  and therefore $\dot \omega = 0$, $\dot p^G_j = 0$.
Hence from the system dynamics, we deduce that the largest invariant set in $\mathcal{S}_1$ for which $\dot V_{1,e}=0$ is the set of equilibrium points in $\mathcal{S}_1$.
Therefore, by LaSalle’s Theorem we have convergence to the set of equilibrium points in $\mathcal{S}_1$. Convergence to an equilibrium point in $\mathcal{S}_1$ for $\epsilon_1$ sufficiently small, can be deduced using arguments analogous to those in \cite{haddad2008nonlinear}[Prop. 4.7, Thm.
4.20], noting that the equilibrium points in $\mathcal{S}_1$ are Lyapunov stable.

We can also deduce that the convergence to an equilibrium point is exponential. This follows by using the fact that\footnote{It can also be shown that due to this property the equilibrium points $x_e^\ast$ for a given equilibrium frequency $\omega^\ast$ are unique.} $\eta_{ij}=\theta_i-\theta_j$, and by considering the trajectories of $x_e$  where the initial conditions of $\eta$ satisfy this property. These trajectories lie in an invariant subspace of $\mathbb{R}^{|E_e|+|N_e|+1}$ where this property is satisfied.

In this subspace, it can be shown that linearization of \eqref{eq.edynamics} about $x_e^\ast$ leads to a linear system where the origin is asymptotically stable, therefore local exponential convergence can be deduced for the nonlinear system in this subspace (using arguments analogous to those in the proof of \cite{khalil2002nonlinear}[Thm.~4.7]).

We will now prove that the average temperature $\bar T$ converges to a constant value. The average temperature dynamics are given by
\begin{align}\dot {\bar T} &=  \frac{1}{\mathcal{V}}\textbf{1}^T  V \dot T
=\frac{1}{\mathcal{V}}\textbf{1}^T [-A_h+h]=\frac{1}{\mathcal{V}}\textbf{1}^T h,
\label{eq:barT}
\end{align}
where
\vspace{-.2cm}
\begin{gather}\label{eq:h}
h=\left[\begin{array}{c} h^G+h^P-h^L \\ \textbf{0} \end{array}\right].
\end{gather}
In \eqref{eq:barT} have used the property $\textbf{1}^TA_h=0$. Variable $h^G$ satisfies the Ordinary Differential Equation (ODE) \eqref{eq.controlh} and $h^P$ equation \eqref{eq.pumpm1}. We will first show that when $h^P$ is constant the corresponding equilibrium point $(\bar T^\ast, h^{G\ast})$ of the state vector $\bar x_h=(\bar T, h^{G})$ is asymptotically stable.

We consider the storage function $V_{1,t}(\bar x_h) =  \frac{1}{2} \mathcal{V} (\bar T - \bar T^*)^2$.
Using the expression for $\dot {\bar T}$ in \eqref{eq:barT} we obtain
\begin{align}
 \dot V_{1,t} & =  \mathcal{V} (\bar T - \bar T^*) \frac{\textbf{1}^T }{\mathcal{V}} V \dot {T} = \sum_{j \in E^G_h} (\bar T - \bar T^*)(h^G_j - h^{G*}_j). \nonumber
\end{align}
We consider $V_{1,s} = \frac{1}{2} \sum_{j \in E_h^G} \frac{1}{\tilde Q_{h,jj}} (h^G_j - h^{G*}_j)^2$, then
\begin{align}
&\dot V_{1,s} =   \sum_{j \in E_h^G}  -\frac{1}{\tilde Q_{h,jj}}(h^G_j - h^{G*}_j)^2 -  (h^G_j - h^{G*}_j)(\bar T - \bar T^*). \nonumber
\end{align}
We then define $V_{1,h}(\bar x_h) = V_{1,t}+ V_{1,s}$ and hence have $\dot V_{1,h}(\bar x_h) = \dot V_{1,t}+ \dot V_{1,s} =  - \sum_{j \in E_h^G} \frac{1}{\tilde Q_{h,jj}} (h^G_j - h^{G*}_j)^2$.

Using $V_{1,h}$ as a Lyapunov-like function we can deduce using Lasalle's theorem that $(\bar T^\ast, h^{G,\ast})$ is asymptotically stable when $h^P$ is constant. Since the system is linear we can deduce that $\bar x_h$ converges exponentially to $(\bar T^\ast, h^{G,\ast})$ also when $h^P$ tends exponentially to its equilibrium value. The latter is the case for $h^P$ from the convergence properties of the frequency $\omega_i$ established in the first part of the proof associated with the electric power system.

We will finally prove that the temperature vector $T$ also converges to a constant value, {given the convergence of average temperature $\bar T$}. We consider deviations of $T$ and variable $h$ in \eqref{eq:h} from equilibrium values $T^\ast$, $h^\ast$, respectively. We denote the deviations as $\tilde T = T-T^\ast$, $\tilde h = h-h^\ast$, and these satisfy the ODE
\begin{gather}\label{eq:tildeT}
\dot{\tilde T}=-V^{-1}A_h\tilde T + V^{-1}\tilde h.
\end{gather}
We make use of the fact that $V^{-1}A_h$ has the same non-zero eigenvalues as $V^{-1/2}A_hV^{-1/2}$.
Due to the fact that $A_h + A_h^T$ is positive semidefinite with a simple eigenvalue at the origin, $V^{-1/2}A_hV^{-1/2}$ also satisfies this property, and we can hence deduce that $V^{-1}A_h$ has all its eigenvalues in the right half-plane with a simple value at the origin.

From our previous analysis associated with the electrical power system and the convergence of $\bar T$ we have that $\tilde h$ converges exponentially to 0 (when the initial condition of the system is sufficiently close to the equilibrium point). Therefore, from the linearity of the system in \eqref{eq:tildeT} and the location of the eigenvalues of $V^{-1}A_h$ we deduce that $\tilde T$ converges to a constant value, and hence $T$ also converges to a constant value, the equilibrium $T^*$. {
{{Hence the} states of the combined heat and power system under Mode~1 converge {to constant} equilibrium values.}} 
This completes the proof.

\subsection{Proof of Theorem \ref{thm:stability} under Mode 2} \label{proof.stability_m2}

{For the} dynamics of the combined heat and power system, we consider the state vector  $\bar x = (\eta, \omega,  p^{G}, \bar T, h^G$), i.e. the heating system is described by its dynamics for the average temperature $\bar T$ with the aggregate system described by \eqref{eq.edynamics}, \eqref{eq.control}, \eqref{eq.pumpmodel}, \eqref{eq.pumpm2}, \eqref{eq:barT}, \eqref{eq:h}.  Below we construct a Lyapunov function for this aggregate system so as to prove convergence of $\bar x$ to an equilibrium point.

We consider storage functions $V_{2,\omega}(\omega) = \frac{1}{2} \sum_{j \in N_e} M_j (\omega_j - \omega^*)^2 $ and $V_{2,l}(\eta) = \sum_{ij \in E_e} B_{ij} \int_{\eta_{ij}^*}^{\eta_{ij}}(\sin \theta  - \sin \eta_{ij}^*) d\theta $ and the time derivative of $V_{2,\omega}(\omega) + V_{2,l}(\eta)$ along the
system trajectories~is
\begin{align}
&\dot V_{2,\omega}
+ \dot V_{2,l}
= \sum_{j \in N_e} -D_j (\omega_j - \omega^*)^2   \\
&~~ +  \sum_{j \in N_e^G} (\omega_j - \omega^* )(p^G_j - p^{G*}_j)  - \sum_{j \in H_e} (\omega_j - \omega^*)(p^P_j - p^{P*}_j). \nonumber
\end{align}

We consider also the storage function $ V_{2,g}(p^G) = \frac{1}{2} \sum_{j \in N_e^G}   \frac{{\tau_{e,j}}}{\tilde Q_{e,jj}}(p^G_j - p^{G*}_j)^2$.
Then
\begin{align}
&\dot V_{2,g} =  \sum_{j \in N_e^G} \frac{1}{\tilde Q_{e, jj}} (p^G_j - p^{G*}_j)^2 -   (p^G_j - p^{G*}_j)(\omega_j - \omega^*) . \nonumber
\end{align}

We consider the storage function $V_{2,t}(\bar T) =  \frac{1}{2} \alpha_i \mathcal{V} (\bar T - \bar T^*)^2$, {where $\alpha_i = \frac{m_i}{C_{o,i}},~i \in E_h$ is a constant coefficient.}

Using
$\dot {\bar T} =  \frac{\textbf{1}^T }{\mathcal{V}} V \dot T $ we obtain
\begin{align}
& \dot V_{2,t}
=  \alpha_i \mathcal{V} \bar T \frac{\textbf{1}^T }{\mathcal{V}} V \dot {\bar T} \nonumber \\
&=  \sum_{j \in H_e}  (\omega_j - \omega^*)(p^P_j - p^{P*}_j) + \sum_{j \in E^G_h} \alpha_i (\bar T - \bar T^*)(h^G_j - h^{G*}_j),  \nonumber
\end{align}
where we have used the property $\textbf{1}^T A_h(T - T^*) = 0$.

We consider also the storage function
\begin{align}
    V_{2,s}(h^G)& = \frac{1}{2} \alpha_i \sum_{j \in E_h^G} \frac{{\tau_{h,j}}}{\tilde Q_{h,jj}}(h^G_j - h^{G*}_j)^2, \nonumber
\end{align}
then
\begin{align}
&\dot V_{2,s}
=   \alpha_i \sum_{j \in E_h^G}  -\frac{1}{\tilde Q_{h,jj}}(h^G_j - h^{G*}_j)^2 - (\bar T - \bar T^*)(h^G_j - h^{G*}_j) . \nonumber
\end{align}

\vspace{-0.4em}
Then consider the candidate Lyapunov function for the aggregate system $V_{2}= V_{2,\omega}+ V_{2,l} +  V_{2,g} + V_{2,t} +  V_{2,s} $:
\begin{align}
&\dot V_2(\bar x) = \dot V_{2,\omega}+ \dot V_{2,l} + \dot V_{2,g} +\dot V_{2,t} + \dot V_{2,s}   \\
&=-\sum_{j \in N_e} D_j  (\omega_j - \omega^*)^2\nonumber \\
&~~~ - \sum_{j \in N_e^G} \frac{1}{\tilde Q_{e,jj}} (p^G_j - p^{G*}_j)^2 -  \sum_{j \in E_h^G} \frac{\alpha_i}{ \tilde Q_{h,jj}} (h^G_j - h^{G*}_j)^2  \nonumber
\end{align}
\vspace{-0.3em}
where equation~\eqref{eq.pumpm2} is used to derive the final expression.

Hence the candidate Lyapunov function $V_{2}(\bar x)$ is non-increasing with time.
From the condition that $\eta_{ij} < \frac{\pi}{2}$, the function $V_2(\bar x)$ has a strict local minimum at $\bar x^* = (\eta^*, \omega^*, p^{G\ast}, \bar T^*, h^{G*}_j)$. Hence we can choose a neighborhood $\mathcal{S}_2$ of $x^*$ given by $\mathcal{S}_2=\{(\eta, \omega, p^{G}, \bar T, h^{G}): V_2 \leq \epsilon_2\}$ for some sufficiently small $\epsilon_2 >0 $ such that this is a compact and {positively}-invariant set that contains $\bar x^*$.

We then apply LaSalle's Theorem, with $V_2$ as the Lyapunov-like function, which states that all trajectories of the system starting from within $\mathcal{S}_2$ converge to the largest invariant set within $\mathcal{S}_2$ that satisfies $\dot V_2(\bar x) = 0$.

 Note that coefficients $D_j$, $\tilde Q_{e,jj}$, and $\tilde Q_{h,jj}$ are all positive, indicating $\dot V_2(\bar x) = 0$  implies  $(\omega_j, p^G_j, h^G_j, \bar T) = (\omega^*, p^{G*}_j, h^{G*}_j, \bar T^*)$ and therefore $\dot \omega = 0$, $\dot p^G_j = 0$,  $\dot {\bar T} = 0$ and $\dot h^G_j = 0$.
Hence from the system dynamics we deduce that the largest invariant set in $\mathcal{S}_2$ for which $\dot V_2=0$ is the set of equilibrium points in $\mathcal{S}_2$.
 Therefore, by LaSalle’s Theorem we have convergence to the set of equilibrium points in $\mathcal{S}_2$.
Convergence to an equilibrium point in $\mathcal{S}_2$ for $\epsilon_2$ sufficiently small, can be deduced using arguments analogous to those in \cite{haddad2008nonlinear}[Prop. 4.7, Thm.
4.20], noting that the equilibrium points in $\mathcal{S}_2$ are Lyapunov stable.

Finally, we can prove the convergence of $T$ to a constant equilibrium $T^*$ from the convergence of $\bar T$ to  $\bar T^*$, following arguments analogous to those used in subsection~\ref{proof.stability_m1}. {{Therefore the states of the combined heat and power system converge to equilibrium values.}}
This completes the proof.

\vspace{-0.1cm}
\subsection{Proof of Proposition~\ref{thm.opti_sep}} \label{proof.opti_sep}
\textbf{Proof:} We will show that at steady state the electrical power system variables are the solution to optimization problem~\eqref{eq.opti_sepe}. Then we will show that at steady state the heating system variables are the solution to the optimization problem~\eqref{eq.opti_seph}.

First, consider the equilibrium of the power network, $\dot {\omega} = 0$ and $\dot p^G = 0$, which results in $p^{G*} = -\tilde{Q}_e \textbf{1} \omega^*$, $p^{P*}_j = a_{1,j} \omega^*,~\forall j \in {H_e}$, and $p^{U*}_j = D_j \omega^*,~\forall j \in N_e$ from the control scheme~\eqref{eq.controle}.

We then consider the optimal solution to the power network optimization problem~\eqref{eq.opti_sepe} which we denote as $(\bar p^{G},\bar p^{P},\bar p^{U})$. Using the KKT conditions for optimality, we obtain
\begin{align}
    Q_e \bar p^G +  \textbf{1} \lambda_1 = 0, ~~
    Q_p \bar p^P -   \textbf{1}\lambda_1 = 0, ~~
    Q_u \bar p^U -   \textbf{1}\lambda_1 = 0,  \nonumber
\end{align}
where $\lambda_1$ is the dual variable associated with constraint~\eqref{eq.opti_ec1}.
Now we set at the optimal point, $\lambda_1 = \omega^*$. Then  $\bar p^G = - \tilde Q_e \textbf{1} \omega^* = p^{G*}$, $\bar p^P =  a_1 \omega^* = p^{P*}$, and $\bar p^U = {D\omega^*} = p^{U*}$, where $a_1$ and $D$ are vectors of $a_{1,j}$ and $D_j$, respectively.  Thus, $(p^{G\ast},p^{P\ast}, p^{U\ast}) = (\bar p^{G},\bar p^{P},\bar p^{U})$.

Given the heat pump heat generation $h^{P*}_{k_j} = C_{o,j} p^{P*}_j$, consider the equilibrium of the heating system, where $\dot {\bar T} = 0$, $\dot T  = 0$, and $h^{G*} = -\tilde{Q}_h \textbf{1} \bar T^*$ from the control scheme~\eqref{eq.controlh}.

Then consider the optimal solution to the {heating} 
network optimization problem~\eqref{eq.opti_seph} which is denoted as $\bar h^G$. Using the KKT conditions we obtain
$Q_h \bar h^G +  \textbf{1} \mu_1 = 0$
where $\mu_1$ is the dual variable associated with constraint~\eqref{eq.opti_hc1}.
Now we {set} at the optimal point, $\mu_1 = \bar T^*$. Then $\bar h^G = - \tilde Q_h \textbf{1} \bar T^* = h^{G*}$. This completes the proof.

\subsection{Proof of Theorem~\ref{thm.opti_joint}} \label{proof.opti_joint}

\textbf{Proof:} We will show that at steady state the system variables are the optimal solution to the combined heat and power optimization problem~\eqref{eq.opti_joint}.

First, consider the equilibrium of the aggregate dynamical system.
At equilibrium, $\dot \omega = 0$, $\dot {\bar T} = 0$,  $\dot {T} = 0$.  Hence $p^G = -\tilde Q_e \textbf{1} \omega^*$ and $h^G = -\tilde Q_h \textbf{1} \bar T^*$ from the control scheme~\eqref{eq.control} and equation~\eqref{eq.pumpm2}.

Then we consider the optimal solution to the combined heat and power optimization problem~\eqref{eq.opti_joint} which we denote as $(\bar p^{G}, \bar h^G, \bar p^{P},\bar p^{U})$. Consider the Lagrangian
for~\eqref{eq.opti_joint}
\begin{align}
&L(p^G, h^G, p^P, \lambda_2, \mu_2) = \frac{1}{2} (p^G)^T Q_e p^G + \frac{1}{2} (h^G)^T \alpha_j Q_h h^G \nonumber \\
&~~+ \frac{1}{2} (p^U)^T Q_u p^U + \mu_2 \left[ \textbf{1}^T h^G - \textbf{1}^T h^L +  \textbf{1}^T C_{o,j} \cdot p^P \right]  \nonumber \\
&~~ + \lambda_2 \left[ \textbf{1}^T p^G - \textbf{1}^T p^L - \textbf{1}^T p^P - \textbf{1}^Tp^U                 \right],  \nonumber
\end{align}
where $\lambda_2$ and $\mu_2$ are dual variables associated with the constraints~\eqref{eq.opti_jointc1} and~\eqref{eq.opti_jointc2}, respectively. Then from the KKT conditions we have
\begin{subequations}
    \begin{align}
    \frac{\partial L}{\partial p^G}  &= Q_e p^G +  \textbf{1} \lambda_2 = 0, \label{eq.ac1}\\
    \frac{\partial L}{\partial h^G}  &= \frac{m_j}{C_{o,j}} \cdot Q_h h^G +  \textbf{1} \mu_2     = 0, \label{eq.ac2}
    \end{align}
    \begin{align}
    \frac{\partial L}{\partial p^P_j}  &= -\lambda_2 + C_{o,j} \cdot \mu_2  = 0, ~~j \in H_e\label{eq.ac3}\\
    \frac{\partial L}{\partial p^U}  &= Q_u\bar p^U - \textbf{1}\lambda_2    = 0, \label{eq.ac5}\\
    \frac{\partial L}{\partial \lambda_2}  &= \textbf{1}^T (p^G - p^L - p^P - p^U ) = 0, \label{eq.ac4} \\
    \frac{\partial L}{\partial \mu_2}      &= \textbf{1}^T (h^G - h^L + C_{o,j} \cdot p^P) = 0.  \label{eq.ac6}
    \end{align}
\end{subequations}

Then we set
$\lambda_2 = \omega^*$. Hence $\mu_2 = \frac{1}{C_{o,j}} \lambda_2 = \frac{1}{C_{o,j}} \omega^*$
from \eqref{eq.ac3}.  Then we have $\bar p^G = - \tilde Q_e \textbf{1} \omega^* = p^{G*}$,  $\bar p^U = D \omega^* = p^{U*}$, and $ \frac{1}{C_{o,j}}\bar h^G = - \frac{1}{C_{o,j}} \tilde Q_h \textbf{1} \frac{\omega^*}{m_j} = \frac{1}{C_{o,j}} h^{G*}, ~ j \in N_e$.
Thus, $(p^{G\ast},h^{G\ast}, p^{U\ast}) = (\bar p^{G},\bar h^{G},\bar p^{U})$. Then from the power balance constraint~\eqref{eq.opti_jointc1}, we know $p^{P\ast} = \bar p^{P}$, thus completing the proof.

\subsection{Proof of Corollary~\ref{thm.general_pas}} \label{proof.general_pas}
In Mode 1, consider a candidate Lyapunov function for the power network, $\hat V_{1,e}(x_e)= V_{1,\omega} + V_{1,l}+ \sum_{j \in N^G_e} V_{e,j}^g(x_{e,j})$. Hence we have
\begin{align}
    & \dot {\hat V}_{1,e}(x_e) =
     \dot V_{1,\omega} + \dot V_{1,l}+ \sum_{j \in N^G_e} \dot V_{e,j}^g(x_{e,j}) \nonumber \\
    &\leq \sum_{j \in N_e} -(D_j +a_{1,j}) (\omega_j - \omega^*)^2 - \sum_{j \in N^G_e} \phi_{e,j}(-\omega_j + \omega^*) \leq 0. \nonumber
\end{align}
Similarly, consider the following storage function for the heating network, $\hat V_{1,h}(x_h) = V_{1,t}+ \sum_{j \in E^G_h} V_{h,j}^g(x_{h,j})$. Hence we have
\begin{align}
\dot {\hat{V}}_{1,h}(x_h) & = \dot V_{1,t}  + \sum_{j \in E^G_h} {\dot V_{h,j}^g} 
=  - \sum_{j\in E_h^G} \phi_{h,j}(-\bar T + \bar T^*) \leq 0. \nonumber
\end{align}
The proof then proceeds in a way analogous to that followed in Appendix~\ref{proof.stability_m1}.

In Mode 2, consider the candidate Lyapunov function for the aggregate system $\hat V_{2}= V_{2,\omega}+ V_{2,l} +  \sum_{j \in N^G_e} V_{e,j}^g(x_{e,j}) + V_{2,t} +  \alpha_j \sum_{j\in E_h^G} V_{h,j}^s $. Hence we have
\begin{align}
\dot {\hat{V}}_2(\bar x)
\leq& \sum_{j \in N_e} -D_j (\omega_j - \omega^*)^2 \nonumber \\
& - \alpha_j \sum_{j\in E_h^G} \phi_{h,j}(-\bar T + \bar T^*) - \sum_{j\in N_e^G} \phi_{e,j}(-\omega_j + \omega^*)  \nonumber
\end{align}
The proof then proceeds in a way analogous to that in Appendix~\ref{proof.stability_m2}.


\bibliographystyle{IEEEtran}
\bibliography{reference}             

\end{document}